\documentclass[journal]{IEEEtran}
\normalsize

\ifCLASSINFOpdf
\else
\fi
\usepackage[dvipsnames]{xcolor}

\usepackage{amssymb,eqnarray,amsmath,stfloats}
\usepackage{graphicx,multirow}

\usepackage[font=small,skip=1pt]{caption}

\usepackage{epstopdf}
\usepackage{setspace}
\usepackage{dsfont}
\usepackage{epsfig}
\usepackage{fancybox}
\usepackage{indentfirst}
\usepackage{extarrows}
\usepackage{caption}
\usepackage{subcaption}
\usepackage{algorithm}
\usepackage{algpseudocode}
\usepackage{color}

\usepackage{multirow}
\usepackage{tabularx}
\usepackage{longtable}
\usepackage{cite}
\usepackage{amssymb}
\usepackage{amsmath}
\usepackage{amsthm}
\usepackage{float}
\usepackage{caption}
\usepackage{slashbox}
\usepackage{graphicx}
\usepackage{float}
\floatstyle{plaintop}
\restylefloat{table}

\makeatother

\allowdisplaybreaks
\graphicspath{{Figures/}}

\begin{document}
 \begin{titlepage}
	\vspace{2 cm}
	This article has been accepted for publication by IEEE.

	\textcopyright 2018 IEEE. Personal use of this material is permitted. Permission
	from IEEE must be obtained for all other uses, in any current or future
	media, including reprinting/republishing this material for advertising or
	promotional purposes, creating new collective works, for resale or
	redistribution to servers or lists, or reuse of any copyrighted
	component of this work in other works.
\end{titlepage}
%
\title{Partial Zero-Forcing for Multi-Way Relay Networks}
%
%
%

\author{Samira~Rahimian, Wuhua~Zhang,    Moslem~Noori, Yindi~Jing,
        and~Masoud~Ardakani
        \thanks{S. Rahimian, Y. Jing, and M. Ardakani are with the Department
                of Electrical and Computer Engineering, University of Alberta, Edmonton,
                AB, T6G 2V4, Canada (e-mail: \{srahimia, yindi, ardakani\}@ualberta.ca).  W. Zhang and M. Noori were with the same department of University of Alberta (e-mail: \{wuhua, moslem\}@ualberta.ca).}
        
               \thanks{Part of the material in this paper was presented at IEEE 16th International Workshop on Signal Processing Advances in Wireless Communications (SPAWC) \cite{zhang2015partial}. 
                }
				\thanks{Manuscript was received October 10, 2017, revised February 11, 2018, and accepted April 7, 2018.}
            }

\maketitle

\begin{abstract}
        The ever increasing demands for mobile network access have resulted in a significant increase in bandwidth usage. By improving the system spectral efficiency, multi-way relay networks (MWRNs) provide promising approaches to address this challenge. In this paper, we propose a novel linear beamforming design, namely partial zero-forcing (PZF), for MWRNs with a multiple-input-multiple-output (MIMO) relay. Compared to zero-forcing (ZF),  PZF relaxes the constraints on the relay beamforming matrix such that only partial user-interference, instead of all, is canceled at the relay. The users eliminate the remaining interferences through self-interference and successive interference cancellation. A sum-rate maximization problem is formulated and solved to exploit the extra degrees-of-freedom resulted from PZF. Simulation results show that the proposed PZF relay beamforming design achieves significantly higher network sum-rates than the existing linear beamforming designs.
\end{abstract}


%
\IEEEpeerreviewmaketitle

\section{Introduction}
%
%
%
%
\IEEEPARstart{T}{he} increasing demands for higher data rates along with the limited bandwidth resources have made the design of bandwidth-efficient communication schemes vital for the future of wireless systems. In recent years, a configuration called multi-way relay networks (MWRNs) \cite{gunduz2013multiway} has been proposed to address this challenge. In an MWRN, multiple users exchange information with the help of one cooperative relay node. By smartly leveraging user-interference, instead of completely avoiding it, MWRNs are able to achieve significantly improved spectral efficiency in wireless communication systems \cite{ong2011capacity}.  Possible applications of MWRNs cover a broad range from cellular communications to wireless sensor networks and satellite communications \cite{ong2010optimal}. 

Early studies in MWRNs are mainly on networks with a single-antenna relay \cite{gunduz2013multiway,ong2010optimal,ong2011capacity,noori2012capacity,noori2012optimal}. For instance, Gunduz \emph{et al.} \cite{gunduz2013multiway} provide upper bounds on the common rate of symmetric Gaussian single-antenna MWRNs and calculate the achievable symmetric rate for amplify-and-forward (AF), decode-and-forward (DF), and compress-and-forward (CF) relaying protocols. In addition, some studies, e.g. \cite{ong2010optimal} and \cite{noori2012optimal}, focus on improving the achievable data rates of MWRNs with a single-antenna relay through suggesting new relaying approaches and scheduling the users' transmission order. 

	The performance of MWRNs can be further improved by employing multiple antennas at the relay \cite{amah2011non1, amah2009non, amah2011non, amarasuriya2013multi, gao2014distributed,degenhardt2013non, li2017joint}. In \cite{amah2011non1} and \cite{amah2009non}, for three different relaying scenarios, called unicasting, multicasting, and hybrid uni/multicasting, linear relay transceive beamforming designs based on zero-forcing (ZF), minimum-mean-square-error (MMSE), and matched filter (MF) are proposed. In another study, the situation when the channel state information (CSI) is not available at the relay is investigated \cite{amah2011non}. For this case, the authors use space-time analog network coding transmission for stationary channels and repetition transmission strategy for non-stationary channels. Another relaying scenario, namely superimposed uni/multicasting, is reported in \cite{degenhardt2013non}, which efficiently combines the MMSE beamforming at the relay with joint receive processing at the users. More specifically, by carefully designing the selection of uni/multicast signals at the relay and the interference cancellation order at the users, the proposed strategy improves the system sum-rate. The authors in \cite{li2017joint}, have considered a MIMO MWRN and designed joint relay beamforming and receiver processing matrices to maximize the minimum received signal-to-interference-plus-noise-ratio (SINR) at the users. For the receiver processing, maximum-ratio-combining (MRC), and ZF are considered. However, the proposed iterative algorithm can have high computational complexity, especially when there exists a large number of users and/or antennas.

In this paper, similar to \cite{amah2011non1} and \cite{amah2009non}, we consider MWRNs with beamforming design at the multi-antenna relay. Following the same model as \cite{amah2011non1,amah2009non,amarasuriya2013multi, gao2014distributed,amah2011non} each user individually decodes only the information that is intended for that user. More specifically, we consider a non-regenerative MWRN where a half-duplex relay equipped with $M$ antennas helps $N$ single-antenna users to receive information from each other. The goal of our work is to maximize the achievable sum-rate of the users. To this end, we introduce a novel idea, named partial zero-forcing (PZF). Unlike ZF relay beamforming, where in each relay broadcasting (BC) transmission phase the interference from all interfering users is forced to be zero [6], our proposed PZF only forces partial interference (the interference from a carefully designed subset of the interfering users) to be zero. Thus PZF allows more degrees-of-freedom in the relay beamforming design. Combined with self-interference cancellation and successive interference cancellation at the users, the proposed PZF relay beamforming allows each user in the MWRN to obtain interference-free observations of information from all other users.

Based on the PZF idea, we formulate the sum-rate maximization problem for the MWRN, which is a constrained multi-dimensional non-linear optimization problem. A numerical method, called modified gradient-ascent method, is proposed to find a joint solution of the PZF relay beamforming matrices for all broadcasting time slots. In addition, to reduce the computational complexity, we propose another method to separately optimize the relay beamforming matrix corresponding to each BC time slot. Simulation results show that the proposed PZF relay beamforming design achieves significantly higher network sum-rate than the existing ZF, MMSE, and MF beamforming designs in \cite{amah2011non1} and \cite{amah2009non}. For example, for a homogeneous 3-user MWRN, we report between 14\% to 200\% sum-rate improvements comparing to ZF, MMSE, and MF schemes. 
  In comparison to \cite{li2017joint}, we report slightly lower sum-rates, but it should be noted that our system models are different. Unlike \cite{li2017joint}, we do not allow joint information decoding at the users or joint relay beamforming and receiver processing. Although, these amendments can improve the performance, this improvement comes at a high computational complexity cost, and its significantly higher processing requirement at the users and in the beamforming optimization stage, can make it less attractive for most of the applications. The interesting observation in our work is that with a relatively simple PZF beamforming a significant sum-rate gain can be achieved.

In this paper, bold upper case letters and bold lower case letters are used to denote matrices and vectors, respectively. For a matrix $\textbf{A}$, its transpose, conjugate, Hermitian, inverse, Moore-Penrose pseudoinverse and trace are denoted by $\textbf{A}^T$, $\textbf{A}^*$, $\textbf{A}^H$, $\textbf{A}^{-1}$, $\textbf{A}^+$, and $\rm tr \left\{\textbf{A}\right\}$, respectively. $\textbf{I}_N$ is the $N\times N$ identity matrix and $\rm diag$$\left\{a_1,\cdots,a_N\right\}$ is a diagonal matrix whose diagonal entries starting from the upper left corner are $a_1,\cdots,a_N$. For a vector $\textbf{a}$, $|\textbf{a}|$ denotes its Euclidean norm, and, finally, $\text{mod}_N{(x)}$ is equal to $x$ modulo $N$.  

\section{System Model} \label{Sec System Model}
The system model of MWRNs includes two parts: the network model and the transceiver protocol, which will be elaborated in the following two subsections.

\subsection{Network Model}
We consider an MWRN consisting of $N$ users (called $u_1,u_2,\cdots,u_N$) and one relay. Each user is equipped with one antenna, while the relay is equipped with $M$ antennas. We assume that $M\ge N$, for the relay to have enough degrees-of-freedom to cancel user interferences. This assumption was also used in \cite{amah2011non1} and \cite{amah2009non}. The extension to the case of $M=N-1$ will be considered in Section \ref{secV}.

Both the users and the relay operate in the half-duplex mode. There are no direct channels among the users and only the channels between the relay and the users are available. The users communicate with each other with the help of the relay. 

Let $\textbf{h}_i=(h_{1,i},h_{2,i},...,h_{M,i})^T$ for $i=1,2,...,N$ be the channel vector between $u_i$ and the relay. Thus $\textbf{H}=[\textbf{h}_1,\textbf{h}_2, ..., \textbf{h}_N]$ is the $M\times N$ channel matrix between all users and the relay. The channels are assumed to follow independent frequency-flat Rayleigh fading, where $h_{m,i}$ follows $\mathcal{C}\mathcal{N}(0,\sigma_{i}^2)$, the circularly symmetric complex Gaussian distribution whose mean is zero and whose variance is $\sigma_{i}^2$. With this, we  imply that the channels between a user and the relay's different antennas have the same variance, while the channels between different users and the relay's antennas can have different variances. Moreover, the channels are assumed to be reciprocal and keep unchanged in each communication block of $N$ time slots.

\subsection{Communication Protocol}
\begin{figure}[!ht]
        \centering
        \begin{subfigure}[b]{0.24\textwidth}
                \centering
                \includegraphics[width=\textwidth]{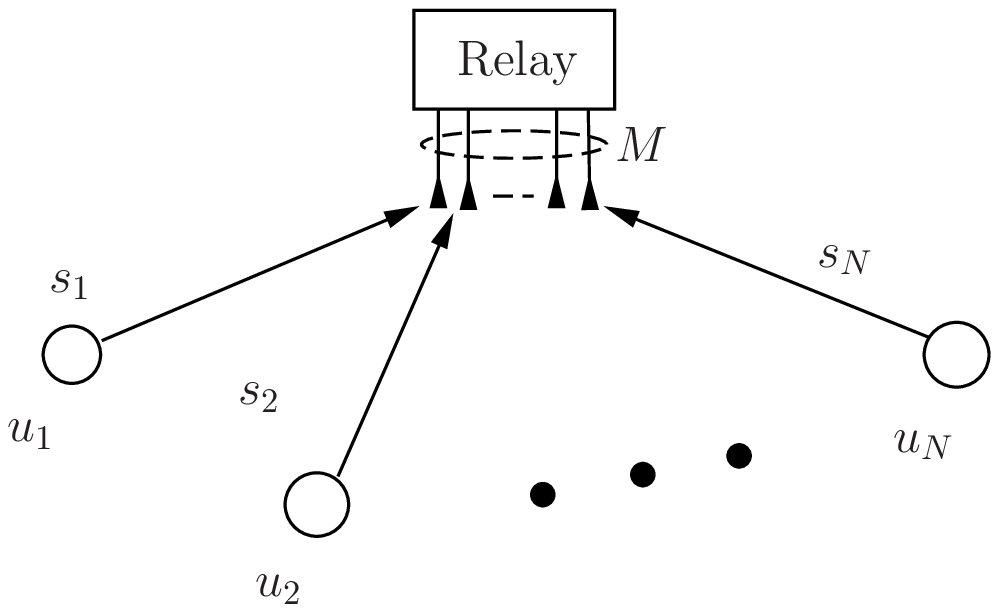}
                \caption{MAC phase.}
                \label{mac}
        \end{subfigure}
        \hfill
        \begin{subfigure}[b]{0.24\textwidth}
                \centering
                \includegraphics[width=\textwidth]{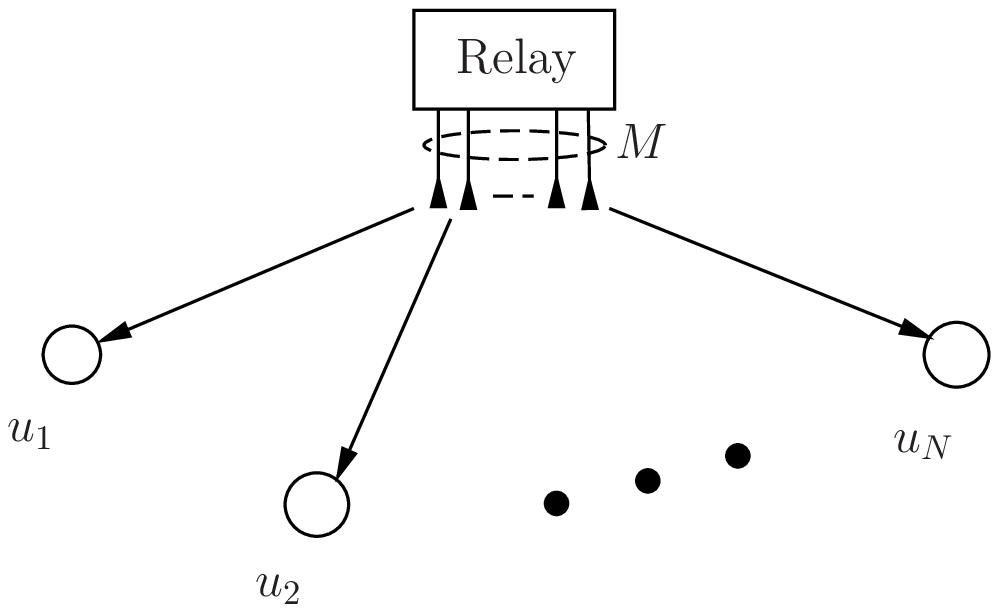}
                \caption{BC phase.}
                \label{bc}
        \end{subfigure}
        \caption{Transceiver protocol of the MWRN.}
        \label{fig:three graphs}
\end{figure}

For all users to send one symbol each to all other users, $N$ time slots are needed, containing $1$ multiple-access (MAC) time slot and $N-1$ BC time slots. In the MAC phase, as shown in Figure \ref{mac}, all users transmit their information symbols simultaneously to the relay. The $M\times 1$ received signal vector at the relay, $\textbf{r}_{\text {RS}}$, is
\begin{equation}
\textbf{r}_{\text {RS}} = \textbf{H}\textbf{s}+\textbf{z}_{\text {RS}}, \label{rr}  \end{equation}
where $\textbf{s}=(s_1,s_2,...,s_N)^T$ is the vector of information symbols of the $N$ users and $\textbf{z}_{\text{RS}}$ is the noise vector at the relay. The transmit power of $u_i$ is denoted as $P_{i}$. Independent Gaussian codebook is used, where the information symbols are assumed to be independent and follow $\mathcal{C}\mathcal{N}(0,P_{i})$. 

In the BC time slots, as shown in Figure \ref{bc}, the multi-antenna relay applies linear beamforming to its received signal vector $\textbf{r}_{\text {RS}}$ and broadcasts information to all users. For the $n$-th BC time slot where $n=1,\cdots,N-1$, $\textbf{G}^{(n)}$ denotes the $M\times M$ relay beamforming matrix. Each user sees the symbols transmitted by the relay other than its intended one as interferences. The symbol transmitted from the relay to each user is changed in every BC time slot, such that after the $N-1$ BC time slots, each user receives the information from all other users. In this section, for the simplicity of presentation, unicasting transmission \cite{amah2011non1} is assumed, where in every BC time slot, the relay transmits different information symbols to different users. Each symbol is intended only for one receiving user in each BC time slot. The extension to hybrid uni/multicasting will be explained in Section \ref{section4}.

\begin{figure}[!ht]
        \centering
        \includegraphics[width=.34\textwidth]{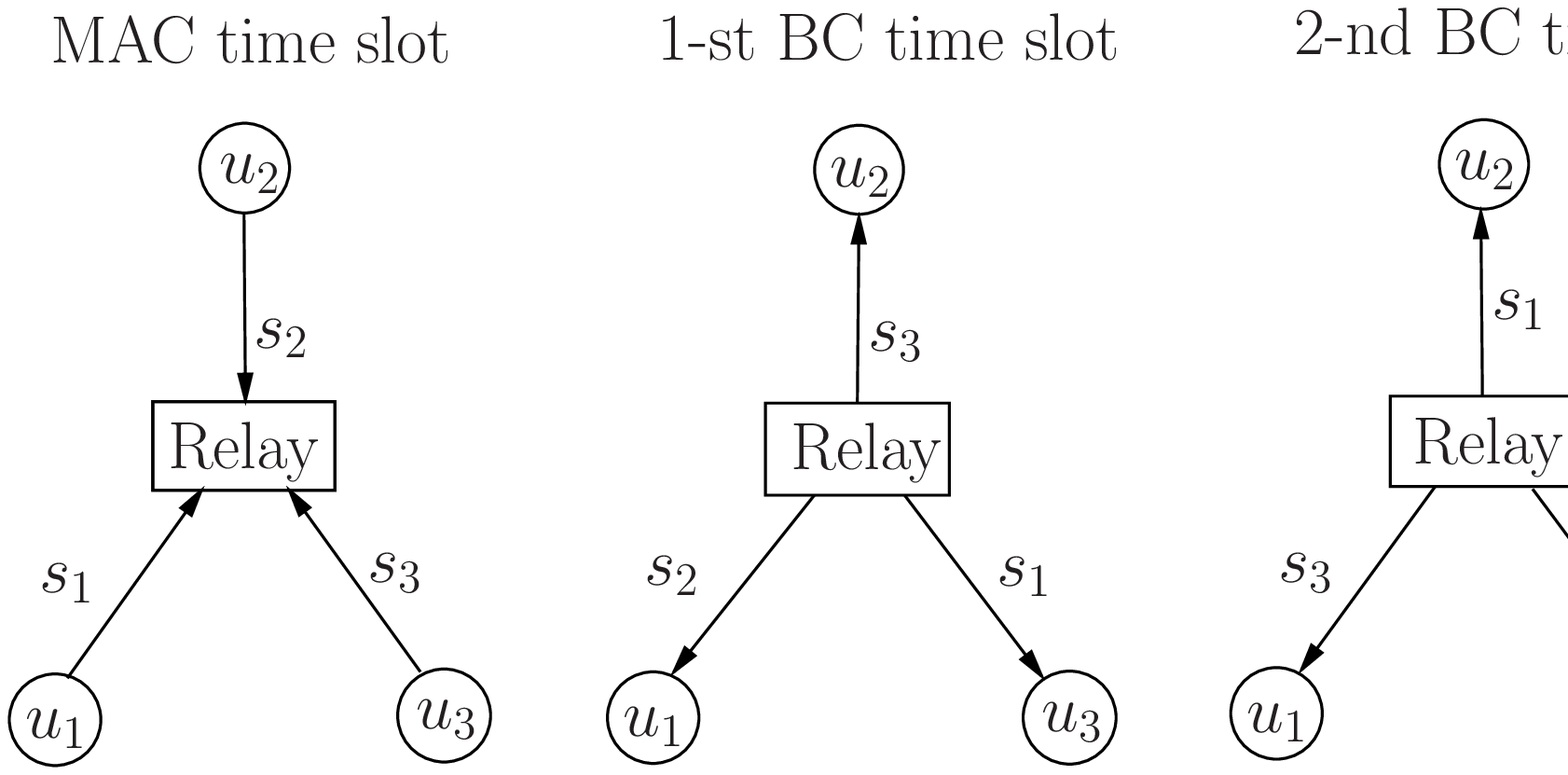}
        \caption{A 3-user MWRN with unicasting strategy.}
        \label{UniModel}
        \vspace{-1.5em}
\end{figure}

To better illustrate the protocol, a 3-user MWRN using unicasting is shown in Figure \ref{UniModel}. In the MAC phase, $u_1$ sends $s_1$, $u_2$ sends $s_2$ and $u_3$ sends $s_3$ simultaneously to the relay. In the first time slot of the BC phase, $u_1$ decodes $s_2$, $u_2$ decodes $s_3$ and $u_3$ decodes $s_1$ from the relay broadcast signal. In the second time slot of the BC phase, $u_1$ decodes $s_3$, $u_2$ decodes $s_1$ and $u_3$ decodes $s_2$. After the MAC phase and the BC phase, each user decodes the information symbols from all other users.

Now, we go back to the general $N$-user MWRNs and explain the BC phase model and the system sum-rate. Because the channels are reciprocal and stationary, the BC channel matrix from the relay to the users is the transpose of the MAC phase channel matrix \textbf{H}. By using (\ref{rr}), the received signal vector of all users in the $n$-th BC time slot, $\textbf{r}_{\text{users}}^{(n)}$, can be written as
\begin{equation}
\textbf{r}_{\text{users}}^{(n)} = \textbf{H}^T\textbf{G}^{(n)}\textbf{H}\textbf{s}+\textbf{H}^T\textbf{G}^{(n)}\textbf{z}_{\text{RS}}+\textbf{z}_{\text{users}}^{(n)},
\label{rr1}
\end{equation}
where $\textbf{z}_{\text{users}}^{(n)}=\left(z_{1}^{(n)},...,z_{N}^{(n)}\right)^T$ is the noise vector at the users in the $n$-th BC time slot. The additive noises at the relay and the users are modeled as independent circularly symmetric complex Gaussian random variables with zero-mean and unit variance, i.e., $\mathcal{CN}(0,1)$. 

The transmit power of the relay for each BC time slot is
\begin{equation}
P_{\text{R}}=\mathbb{E}\{\text{tr}\{\textbf{G}^{(n)}(\textbf{H}\textbf{s}+\textbf{z}_{\text{RS}})[\textbf{G}^{(n)}(\textbf{H}\textbf{s}+\textbf{z}_{\text{RS}})]^H \} \}.
\end{equation}
After straightforward calculations, it can be simplified as
\begin{equation}
P_{\text R}=\text{tr}\Big\{\textbf{G}^{(n)}\big(\textbf{H}\textbf{P}_{\textbf{s}}\textbf{H}^H+\textbf{I}_{M}\big)\big(\textbf{G}^{(n)}\big)^H \Big\},
\label{PowerCons}
\end{equation}
where $\textbf{P}_{\textbf{s}}=\rm {diag}$$\left\{P_1, P_2, ..., P_N\right\}$.

After receiving the relay's signal in the $n$-th BC time slot, $u_k$ decodes $u_i$'s information symbol, which is $s_i$. In this work, the order of decoding is designed as the following relation among $i$, $k$, and $n$
\begin{equation} \label{relation}
i=\text{mod}_N(k+n-1)+1.
\end{equation}
Accordingly, from (\ref{rr1}), the received signal at $u_k$ in the $n$-th BC time slot can be written as
\begin{equation}
r_{k}^{(n)} = \textbf{h}_k^T \textbf{G}^{(n)} \textbf{h}_i s_i +\hspace{-2mm}
\sum^{N}_{j=1,j\neq i}\textbf{h}_k^T\textbf{G}^{(n)}\textbf{h}_js_j+\textbf{h}_k^T\textbf{G}^{(n)}\textbf{z}_{RS}+z_{k}^{(n)}.
\label{4}
\end{equation}
Notice that in each BC time slot, the signal transmitted by the relay contains signals of all users sent in the MAC phase. In (\ref{4}), the first term contains the intended signal from $u_i$, the second term contains the interferences from other users than the intended user (including the receiver's own signal $s_k$), which are all forwarded to the user by the relay, the third term contains the noise propagated from the relay, and the last term is the noise at $u_k$. Thus, the SINR for the communication from $u_i$ to $u_k$, denoted as $\gamma_{k,i}$, can be calculated to be
\begin{equation}\label{SINR_NC}
\gamma_{k,i}=\frac{P_i|\textbf{h}_k^T \textbf{G}^{(n)} \textbf{h}_i|^2}{\sum^{N}_{j=1, j\neq i}P_j|\textbf{h}_k^T \textbf{G}^{(n)} \textbf{h}_j|^2+|\textbf{h}_k^T \textbf{G}^{(n)}|^2+1}.
\end{equation}
\allowdisplaybreaks
However, after each BC time slot, $u_k$ performs interference cancellation by subtracting its self-interference and the interference of users' symbols which have already been decoded in the previous BC time slots, thus the SINR after interference cancellation is
\begin{equation}\label{SINR_C}
\gamma_{k,i}=\frac{P_i|\textbf{h}_k^T \textbf{G}^{(n)} \textbf{h}_i|^2}{\sum^{N}_{j=1, j\neq i, j\neq k, j \notin \mathbb{L}_{k,n}}P_j|\textbf{h}_k^T \textbf{G}^{(n)} \textbf{h}_j|^2+|\textbf{h}_k^T \textbf{G}^{(n)}|^2+1},
\end{equation}
where $\mathbb{L}_{k,n}=\{\text{mod}_N(k+q-1)+1,q = 1,2,...,n-1\}$. $\mathbb{L}_{k,n}$ contains the indexes of the symbols already decoded by $u_k$ from previous $n-1$ BC time slots which is determined by the order of detection defined in \eqref{relation}.
\allowdisplaybreaks
The achievable rate from $u_i$ to $u_k$, denoted as $R_{k,i}$ is thus
\begin{equation}\label{Rate}
R_{k,i}=\log_{2}(1+\gamma_{k,i}).
\end{equation}
The common rate $R_i$ that $u_i$ can reliably send to all other users is
\begin{equation}\label{MinRate}
R_i=\min_{k\ne i}R_{k,i}.
\end{equation}
The achievable sum-rate of the MWRN is thus \cite{amah2011non1},
\begin{equation}\label{SumRate}
R_{\rm sum}=\frac{N-1}{N}\sum^{N}_{i=1}R_i.
\end{equation}

We continue this section by introducing the existing beamforming strategies in the following. 
\subsection{Existing Beamforming Designs}
The sum-rate of MWRNs is given by (\ref{SINR_NC})-(\ref{SumRate}). It is conceivable that the design of relay beamforming matrices, $\textbf{G}^{(1)}$,$\textbf{G}^{(2)}$,$\cdots$ $\textbf{G}^{(N-1)}$, is crucial for the sum-rate performance. In this section, a brief review of existing relay beamforming designs including ZF, MMSE, and MF proposed in \cite{amah2011non1,amah2010beamforming}, is given. It should be noticed that the relay beamforming schemes for MWRNs serve both as receive and transmit beamforming. Hence, they are also called transceive beamforming. In \cite{amah2011non1,amah2010beamforming}, the relay transceive beamforming matrix has the following general structure
\begin{equation}
\textbf{G}^{(n)}=\textbf{G}_{\text{TX}}^{(n)}\textbf{P}^{n}\textbf{G}_{\text{RX}},
\label{GZF}
\end{equation}
where $\textbf{P}$ is the permutation matrix, obtained by circularly shifting the columns of $\textbf{I}_N$ one position to the right. $\textbf{P}^{n}$ is thus the permutation matrix to define the relationship between an arbitrary receiving user, $u_k$, and the corresponding transmitting user, $u_i$, in the BC time slot, $n$. $\textbf{G}_{\text{RX}}$ is the receive beamforming matrix and $\textbf{G}_{\text{TX}}^{(n)}$ is the transmit beamforming matrix. The following designs of $\textbf{G}_{\text{RX}}$ and $\textbf{G}_{\text{TX}}^{(n)}$ have been proposed in \cite{amah2011non1,amah2010beamforming}.

\subsubsection{Zero-Forcing Design}
In ZF, $\textbf{G}^{(n)}$ is designed such that the second term in (\ref{4}) equals to 0 for all $n=1,\cdots,N-1$. That is, the interference from all other users except $u_i$, is forced to zero at $u_k$. $\textbf{G}_{\text{RX}}$ and $\textbf{G}_{\text{TX}}^{(n)}$ are defined as follows
\setlength{\arraycolsep}{1pt}
\begin{eqnarray}
&\textbf{G}_{\text{RX}}&=(\textbf{H}^H \textbf{H})^{-1}\textbf{H}^H,\nonumber\\
&\textbf{G}_{\text{TX}}^{(n)}&=\frac{1}{p_{\text{ZF}}^{(n)}}\textbf{H}^* (\textbf{H}^T\textbf{H}^*)^{-1},
\label{RS-E}
\end{eqnarray}
\setlength{\arraycolsep}{5pt}  where $p_{\text{ZF}}^{(n)}$ is used to fulfill the power constraint at the relay.

\subsubsection{Minimum-Mean-Square-Error Design}
MMSE beamforming minimizes the mean square error of the signal. For MWRNs, the MMSE receive and transmit beamforming matrices are
\begin{eqnarray}
&\textbf{G}_{\text{RX}}&\hspace{-2mm}=\textbf{P}_{\textbf{s}}\textbf{H}^H(\textbf{H}\textbf{P}_{\textbf{s}}\textbf{H}^H+\textbf{I}_M)^{-1},\nonumber\\
&\textbf{G}_{\text{TX}}^{(n)}&\hspace{-2mm}=\frac{1}{p_{\text{MMSE}}^{(n)}}(\textbf{H}^*\textbf{H}^T+\frac{N\textbf{I}_M}{P_{\text R}})^{-1}\textbf{H}^*,
\label{eq:MMSE}
\end{eqnarray}
where $p_{\text{MMSE}}^{(n)}$ is used to fulfill the relay power constraint. It is worth mentioning that regularized zero-forcing (RZF) beamforming \cite{peel2005vector}, which is a modification of MMSE, is more practically applicable today. RZF replaces $\textbf{I}_M$ in the MMSE receiver beamforming formula ($\textbf{G}_{\text{RX}}$ in (\ref{eq:MMSE})) with $\alpha \textbf{I}_M$.
\subsubsection{Matched Filter Design}
MF beamforming is the optimal linear beamforming for maximizing the signal-to-noise-ratio (SNR) in the presence of additive noise. The MF receive and transmit beamforming matrices are
\begin{eqnarray}                 &\textbf{G}_{\text{RX}}&\hspace{-2mm}=\textbf{P}_{\textbf{s}}\textbf{H}^H(\textbf{H}\textbf{P}_{\textbf{s}}\textbf{H}^H+\textbf{I}_M)^{-1},\nonumber\\
&\textbf{G}_{\text{TX}}^{(n)}&\hspace{-2mm}=\frac{1}{p_{\text{MF}}^{(n)}}\textbf{H}^*,
\end{eqnarray}
where $p_{\text{MF}}^{(n)}$ is used to fulfill the relay power constraint.

\section{PZF Relay Beam-forming Design} \label{sec PZF}
Based on the ZF relay beamforming design, we propose a new design called PZF. In this section, first we explain the idea of PZF, then we formulate the optimization problem of PZF design for sum-rate maximization. A numerical method called modified gradient-ascent is proposed to solve the optimization problem. Finally, simulation results on the performance of PZF and the comparison with existing beamforming designs are given.

\subsection{PZF Main Idea}
In the ZF relay beamforming design of \cite{amah2011non1}, in all $N-1$ BC time slots, the relay beamforming matrices are designed such that at each user, the effects of transmitted signals of all users except for the desired one are forced to be zero. For instance, if $u_k$ wants to receive $u_i$'s message in the BC time slot, n, all interference signals from $u_j, j \neq i$, (i.e. all terms in \eqref{4} containing $s_j, j \neq i$) are forced to be zero by the relay beamforming matrix $\textbf{G}_{\rm ZF}^{(n)}$. This puts heavy constraints on the relay beamforming matrices $\textbf{G}_{\rm ZF}^{(1)},\cdots,\textbf{G}_{\rm ZF}^{(N-1)}$, i.e., for each $\textbf{G}_{\rm ZF}^{(n)}$, $N(N-1)$ entries of $\textbf{H}^T\textbf{G}_{\rm ZF}^{(n)}\textbf{H}$ must be zero, as can be seen in \eqref{RS-E}. However, such heavy constraints are not necessary to obtain interference-free observations at the users.

 Knowing its own information and the CSI, every user can conduct self-interference cancellation. In addition, up to the $n$-th BC time slot, every user has already decoded the symbols of $n-1$ users, through the previous $n-1$ relay broadcasts, thus it can cancel the interference from these users without further help from the relay. So, the relay beamforming matrix for the $n$-th BC time slot only needs to be designed to cancel the interference from the remaining $N-n-1$ users. This constraint relaxation, which we refer to as PZF, allows more degrees-of-freedom in the design of relay beamforming matrices to improve the network sum-rate. 

In order to better illustrate the PZF design idea and to help later analysis, we define
\begin{equation}
\textbf{A}^{(n)}=\textbf{H}^T\textbf{G}^{(n)}\textbf{H},
\label{AA}
\end{equation}
which as seen from (\ref{rr1}), is the equivalent channel matrix of the $n$-th BC time slot. With ZF, as shown in \eqref{RS-E}, $\textbf{A}^{(n)}$ should be equal to the permutation matrix $\textbf{P}^{(n)}$ where $N(N-1)$ of the entries are zero and $N$ of the entries are 1. However, with PZF only $(N-n-1)N$ entries of $\textbf{A}^{(n)}$ need to be zero and other entries can take any complex number.\footnote{Recall the system equation in (\ref{rr1}), where $\textbf{H}^T\textbf{G}^{(n)}\textbf{H}\textbf{s}=
\textbf{A}^{(n)}\textbf{s}$. Each entry of $\textbf{A}^{(n)}\textbf{s}$ contains $n-1$ previously detected symbols, self-interference, and new symbols that are to be detected in the future. In the proposed PZF beamforming, the idea is to eliminate the interference from symbols to be detected in future via the relay beamforming matrix design and eliminate the interference from previously detected symbols and self-interference via direct interference cancellation at the users. This means that in each row of $\textbf{A}^{(n)}$, our design requires having  $N-(n+1)$ zero entries at predetermined locations, while the rest $n+1$ entries can take any value. So, in total for all the $N$ rows, $\textbf{A}^{(n)}$ matrix should have $(N-n-1)N$ zero entries.}

Take the MWRN where $M=N=3$ for an example. If ZF beamforming is used at the relay, $\textbf{G}^{(1)}$ and $\textbf{G}^{(2)}$ should be designed so that $\textbf{A}^{(1)}$ and $\textbf{A}^{(2)}$ have the following forms
\begin{equation}
\textbf{A}^{(1)}_{\text{ZF}}=
\left(                
\begin{array}{ccc}   
0    & 1      & 0\\  
0 & 0      & 1  \\ 
1      & 0 & 0
\end{array}
\right),
\textbf{A}^{(2)}_{\text{ZF}}=
\left(                
\begin{array}{ccc}   
0    & 0      & 1\\  
1 & 0      & 0  \\ 
0      & 1 & 0
\end{array}
\right).
\end{equation}
Both $\textbf{A}^{(1)}_{\text{ZF}}$ and $\textbf{A}^{(2)}_{\text{ZF}}$ should have 6 zero-value entries, which means all the interference signals except the desired one are canceled through ZF relay beamforming. 

However, if PZF beamforming is used at the relay, $\textbf{A}^{(1)}_{\text{PZF}}$ and $\textbf{A}^{(2)}_{\text{PZF}}$ are supposed to have the following forms
\begin{equation}
\textbf{A}^{(1)}_{\text{PZF}}=
\left(                
\begin{array}{ccc}   
* & * & 0\\  
0 & * & *  \\ 
* & 0 & *
\end{array}
\right),
\textbf{A}^{(2)}_{\text{PZF}}=
\left(                
\begin{array}{ccc}   
* & * & *\\  
* & * & *  \\ 
* & * & *
\end{array}
\right),
\end{equation}
where ``$*$'' means that the entry can take any complex number. This way the restrictions on $\textbf{A}^{(1)}$ and $\textbf{A}^{(2)}$ are reduced. Only 3 entries in $\textbf{A}^{(1)}_{\text{PZF}}$ should be zero and all others can take any complex number. In the first BC time slot, the relay beamforming matrix only needs to be designed to cancel part of the interferences and the rest can be canceled through self-interference cancellation at the users. In the second BC time slot,  the relay leaves the interferences  to be entirely canceled by the users, as the users have the knowledge of their own information symbols and also already decoded information symbols in the first BC time slot.
\vspace{-1pc}

\subsection{PZF Formulation}
In this section, we formulate the PZF beamforming design and specify the relay beamforming matrix optimization problem. 

First, we specify the structure of $\textbf{A}^{(n)}$ for PZF. We denote the $(i,j)$-th element of $\textbf{A}^{(n)}$ as $a_{ij}^{(n)}$. To clearly express the PZF constraints on $\textbf{A}^{(n)}$, a set of 3-tuple indexes are introduced as the following

\begin{equation}
\mathbb{A} =\left\{(i,j,n)
\left| 
\begin{array}{l}
n \hspace{-0.4mm}=1,2,\cdots,N-2; \\ 
i \hspace{0.5mm}=1,2,\cdots,N; \\ 
q =1,2,\cdots,N-n-1; \\
j =\text{mod}_N{(i+q+n-1)+1}
\end{array}\right. \right\}, 
\label{A-set}
\end{equation} which is a subset of the 3-tuple indexes $(i,j,n)$ representing the receiving user, the transmitting/interfering user, and the BC time slot. A 3-tuple index is an element of $\mathbb{A}$, if in the $n$-th BC time slot, the interference of $u_j$ to $u_i$ needs to be canceled under the PZF design.

From \eqref{SINR_C}-\eqref{SumRate}, the sum-rate maximization problem can be stated mathematically, as
\begin{eqnarray}\label{JointOpt0}
&&\hspace{-8mm}\max_{\textbf{G}^{(1)},\cdots,\textbf{G}^{(N-1)}}\hspace{-1mm}\sum^{N}_{i=1}\min_{k \neq i}\left\{\log_{2}\Bigg(1\hspace{-1mm}+\hspace{-1mm}\frac{{P_i|\textbf{h}^T_k\textbf{G}^{(n)}\textbf{h}_i|}^2}{{|\textbf{h}_k^T\textbf{G}^{(n)}|}^2\hspace{-1mm}+1}\Bigg)\hspace{-1mm} \right\} \label{J-opt10}\\
\hspace{-5mm}&\text{s.t.}& {\rm tr}\left\{\textbf{G}^{(n)}\left(\textbf{H}\textbf{P}_{\textbf{s}}\textbf{H}^{H}+\textbf{I}\right)(\textbf{G}^{(n)})^{H}\right\}\le P_{R}, \label{J-opt20}\\
&\mbox{and}& [\textbf{H}^T\textbf{G}^{(n)}\textbf{H}]_{(i, j)}=0, \ \text{for} \ (i, j, n) \in \mathbb{A}. \label{nonCon0}
\label{opt-prob0}
\end{eqnarray}
The non-linear constraint in (\ref{J-opt20}) is due to the transmit power constraint at the relay and the linear constraints in (\ref{opt-prob0}) are forced by the PZF idea.
This sum-rate maximization problem is a multi-dimensional non-linear optimization problem with linear and non-linear constraints. So, first we simplify the problem using transformation. The optimization variables are beamforming matrices $\textbf{G}^{(1)},\cdots,\textbf{G}^{(N-1)}$. After applying the transformation in (\ref{AA}), the problem can be converted to an optimization over $\textbf{A}^{(1)},\cdots,\textbf{A}^{(N-1)}$, and $\textbf{G}^{(n)}$ can be calculated from $\textbf{A}^{(n)}$ using
\begin{equation}
\textbf{G}^{(n)}=\big(\textbf{H}^{T}\big)^{+}\textbf{A}^{(n)}\textbf{H}^{+}.
\label{GG}
\end{equation}
This transformation makes the linear constraints in (\ref{nonCon0}) simpler which in turn simplifies the optimization problem.
Thus, the sum-rate maximization problem is transformed as
\begin{eqnarray}\label{JointOpt}
&&\hspace{-8mm}\max_{\textbf{A}^{(1)},\cdots,\textbf{A}^{(N-1)}}\hspace{-1mm}\sum^{N}_{i=1}\min_{k \neq i}\left\{\log_{2}\Bigg(1\hspace{-1mm}+\hspace{-1mm}\frac{{P_i|\textbf{h}^T_k\textbf{G}^{(n)}\textbf{h}_i|}^2}{{|\textbf{h}_k^T\textbf{G}^{(n)}|}^2\hspace{-1mm}+1}\Bigg)\hspace{-1mm} \right\} \label{J-opt1}\\
\hspace{-5mm}&\text{s.t.}& {\rm tr}\left\{\textbf{G}^{(n)}\left(\textbf{H}\textbf{P}_\textbf{s}\textbf{H}^{H}+\textbf{I}\right)(\textbf{G}^{(n)})^{H}\right\}\le P_{R}, \label{J-opt2}\\
&\mbox{and}& a_{i j}^{(n)}=0, \mbox{for } (i,j,n) \in \mathbb A.
\label{opt-prob}
\end{eqnarray}

\subsection{Joint Optimization of the Relay Beamforming Matrices}\label{sec-joint}
In this subsection, we provide a numerical method to jointly optimize all $\textbf{A}^{(n)}$ matrices. We define
\begin{equation}
\textbf{x}^{(n)}=[a^{(n)}_{11}\hspace{-1mm}\ \ a^{(n)}_{12}\hspace{-1mm}\ \ \cdots \ \ \hspace{-1mm}a^{(n)}_{ij}\hspace{-1mm}\ ((i,j,n) \notin \mathbb A)\hspace{-1mm}\ \ \cdots \ \ \hspace{-1mm}a^{(n)}_{N,N}], \label{xxn1}\\
\end{equation}
which includes all the  nonzero entries in $\textbf{A}^{(n)}$ and is $U_n$-dimensional  where 
\begin{equation}\label{NumNonzero}
U_n=(n+1)N, \text{for} \ \  n=1,2,\cdots,N-1.
\end{equation}
 Further, we define vector $\textbf{x}$ formed by concatenating all the vectors $\textbf{x}^{(n)}$, as
\begin{equation}
\textbf{x}=[\textbf{x}^{(1)}, \textbf{x}^{(2)}, \cdots,\textbf{x}^{(n)}, \cdots, \textbf{x}^{(N-1)}]. \label{xxn2}
\end{equation}
It contains $a_{ij}^{(n)}$s for $(i,j,n)\notin \mathbb A$ and is $W$-dimensional,
\begin{equation}
W=(N+2)N(N-1)/2.
\label{NumOpt}
\end{equation}
With these notations, the optimization problem in (\ref{J-opt1}) to (\ref{opt-prob}) can be written as an optimization problem over $\textbf{x}$ and the constraints in (\ref{opt-prob}) are naturally eliminated. Since the objective function in \eqref{J-opt1} is non-convex and the constraints in (\ref{J-opt2}) are non-linear, the solution is in general difficult to find. A common method to find sub-optimal solutions for such problems is to use the gradient-ascent method. However, the conventional gradient-ascent method does not work well in our case because of the complicated non-linear constraint. Actually, by moving toward the gradient direction even with a small step size, the new $\textbf{x}$ vector may violate the power constraint. To avoid this, we propose a modification to the gradient-ascent method. Our \emph{modified gradient-ascent method} updates the $\textbf{x}$ vector toward the direction of the \textit{modified gradient} specified in what follows.
\subsubsection{Modified Gradient}
Denote the objective function in $(\ref{J-opt1})$ as $f(\textbf{x})$ and the power constraint in (\ref{J-opt2}) as $\phi(\textbf{x}^{(n)})\le P_R$, where
\begin{equation}
\phi\big(\textbf{x}^{(n)}\big)=\phi\big(\textbf{A}^{(n)}\big)={\rm tr}\left\{\textbf{G}^{(n)}\big(\textbf{H}\textbf{P}_\textbf{s}\textbf{H}^{H}+\textbf{I}\big)\big(\textbf{G}^{(n)}\big)^{H}\right\}.
\end{equation}
So, the optimization problem becomes
\begin{eqnarray}
&&\label{f_jontopt} \hspace{-5cm}\max_{\textbf{x}} f(\textbf{x})\\ 
 \mbox{\ \ \ s.t.\ } \phi(\textbf{x}^{(n)})\le P_R 
\mbox{ for } n=1,2,\cdots, N-1.
\label{const_jointopt}
\end{eqnarray}

Notice from the definitions in (\ref{xxn1})-(\ref{xxn2}) that the $m$-th element in $\textbf{x}$ is the $l$-th element of $\textbf{x}^{(n)}$ with the relationship $m=2N+ \cdots +nN+l$. Letting ${\bf e}_l$ be the $l$-th canonical basis vector, we define the power normalization factors as
\begin{equation}
\alpha_m^{\rm Re}=\frac{\phi(\textbf{x}^{(n)}+\epsilon \textbf{e}_l)}{P_R} \mbox{\ and \ }
\alpha_m^{\rm Im}=\frac{\phi(\textbf{x}^{(n)}+i\epsilon \textbf{e}_l)}{P_R}.
\end{equation}
So, the modified partial derivative of $f$ with respect to the $m$-th element of $\textbf{x}$, is given by

\begin{align}
\begin{split}\label{partialD}
d(f,x_m) =&\lim_{\epsilon\to 0} {\frac{f\Big(\textbf{x}^{(1)}, \cdots, \frac{\textbf{x}^{(n)}+\epsilon \textbf{e}_l}{\alpha_{m}^{\rm Re}}, \cdots, \textbf{x}^{(N-1)}\Big)-f(\textbf{x})}{\epsilon}}\\ 
+i&\lim_{\epsilon\to 0} {\frac{f\Big(\textbf{x}^{(1)}, \cdots, \frac{\textbf{x}^{(n)}+i\epsilon \textbf{e}_l}{\alpha_{m}^{\rm Im}}, \cdots, \textbf{x}^{(N-1)}\Big)-f(\textbf{x})}{\epsilon}}.
\end{split}
\end{align}
Compared with the definition of normal partial derivative,
\begin{align}
\begin{split}
\dfrac{\partial f}{\partial x_m} =&\lim_{\epsilon\to 0} {\frac{f\Big(\textbf{x}^{(1)}, \cdots, \textbf{x}^{(n)}+\epsilon \textbf{e}_l, \cdots, \textbf{x}^{(N-1)}\Big)-f(\textbf{x})}{\epsilon}}\\ 
+i&\lim_{\epsilon\to 0} {\frac{f\Big(\textbf{x}^{(1)}, \cdots, \textbf{x}^{(n)}+i\epsilon \textbf{e}_l, \cdots, \textbf{x}^{(N-1)}\Big)-f(\textbf{x})}{\epsilon}},
\end{split}
\end{align}
\eqref{partialD} takes the non-linear constraint $\phi(\textbf{x}^{(n)})\le P_R$ into account. In other words, to make sure that this constraint is not violated when $\textbf{x}^{(n)}$ is modified to $\textbf{x}^{(n)}+\epsilon \textbf{e}_l$ or $\textbf{x}^{(n)}+\epsilon i\textbf{e}_l$, the vector is scaled by $\alpha_m^{\rm Re}$, or $\alpha_m^{\rm Im}$, whose definition guarantees the power constraint. The modified gradient of $f$ is thus,
\begin{equation}
D(f,\textbf{x}) = [d(f,x_1)\ \ \cdots\ \ d(f,x_m)\ \ \cdots\ \ d(f,x_W)].
\end{equation}

\subsubsection{Optimization Algorithm}
In our numerical method, $\textbf{x}$ vector is updated toward the modified gradient with a step size $\alpha$. Also, scaling is done at every iteration to guarantee that each searched point satisfies the constraint. In fact, a new point is found by two moves. First, a move of $\textbf{x}$ proportional to the modified gradient is made. Second, constructed from $\textbf{x}$, $\textbf{A}^{(1)},\cdots,\textbf{A}^{(N-1)}$ are scaled to make the power constraint satisfied. $\textbf{x}$ is then moved to a new point accordingly. Once a solution for $\textbf{x}$ is found, we can reconstruct $\textbf{A}^{(1)},\cdots,\textbf{A}^{(N-1)}$, and then from (\ref{GG}) calculate $\textbf{G}^{(1)},\cdots,\textbf{G}^{(N-1)}$. It should be noted that similar to the gradient-ascent method, the proposed modified gradient-ascent method cannot guarantee the global optimality of the solution. However, we can use ZF relay beamforming matrices as the initial point to guarantee a solution better than ZF. The algorithm is described in Algorithm \ref{JointAlgorithm}.
\begin{algorithm}[!ht]
        \caption{Joint optimization scheme.}
        \label{JointAlgorithm}
        \begin{algorithmic}[1]
                \State Initialize $\alpha,tolerance,\textbf{A}^{(n)}$s, $\textbf{x}$ and calculate $D(f,\textbf{x})$.
                \While {$norm(D(f,\textbf{x}))\geq tolerance$}
                \State Update $\textbf{x}$: $\textbf{x}=\textbf{x}+\alpha D(f,\textbf{x})$.
                \State Construct $\textbf{A}^{(n)}$s from $\textbf{x}$.
                \State Scale $\textbf{A}^{(n)}$s based on the constraint and construct $\textbf{x}$. 
                \State Calculate $D(f,\textbf{x})$.
                \EndWhile
                \State Calculate $\textbf{G}^{(1)},\cdots,\textbf{G}^{(N-1)}$ using (\ref{GG}).
             \end{algorithmic}
\end{algorithm}
\vspace{-1em}
\subsection{Separate Optimization of the Relay Beamforming Matrices}
In the method described in Section \ref{sec-joint}, the matrices $\textbf{A}^{(1)},\cdots,\textbf{A}^{(N-1)}$ are jointly optimized and thus the algorithm can be computationally expensive for large MWRNs. In this section, we propose to use separate optimization where the optimization over $\textbf{A}^{(n)}$s for $n=1,\cdots,N-1$ is conducted separately and sequentially.

Notice that the relay beamforming matrix for the $n$-th BC time slot, $\textbf{G}^{(n)}$, directly affects the transmission rates $R_{k,i}$ during this phase, where $k$, $i$ and $n$ satisfy the relation in (\ref{relation}). It does not affect the transmission rates of previous or later BC time slots if ideal source coding and detection are assumed. Thus, we propose to optimize $\textbf{G}^{(n)}$, or equivalently $\textbf{x}^{(n)}$ by maximizing the sum-rate in the $n$-th BC time slot, given by
\begin{equation}
R^{{(n)}}_{\rm sum}=\sum^{N}_{i=1}\log_{2}\Bigg(1+\frac{{P_i|\textbf{h}^T_k\textbf{G}^{(n)}\textbf{h}_i|}^2}{{|\textbf{h}_k^T\textbf{G}^{(n)}|}^2+1}\Bigg).
\end{equation}
Thus, the optimization problem would be
\begin{eqnarray}
&&\hspace{-5cm}\max_{\textbf{x}^{(n)}}R^{{(n)}}_{\rm sum}\\
 \mbox{\ \ \ s.t.\ }  \phi(\textbf{x}^{(n)})\le P_R \mbox{ for } n=1,2,\cdots, N-1.
\label{P2}
\end{eqnarray}
In solving the above optimization problem, the same modified gradient-ascent method is used.  Considering that the number of constraints on $\textbf{A}^{(n)}$ decreases as $n$ increases, we optimize $\textbf{A}^{(n)}$s sequentially with $\textbf{A}^{(1)}$ being the first and $\textbf{A}^{(N-1)}$ being the last. The algorithm for separate optimization is clarified in Algorithm \ref{SepAlgorithm}.
\begin{algorithm}[!ht]
        \caption{Separate optimization scheme.}\label{SepAlgorithm}
        \begin{algorithmic}[1]
                \State Initialize $\alpha$ and $tolerance$.
                \For {$n=1:N-1$}
                \State Initialize $\textbf{A}^{(n)}$, $\textbf{x}^{(n)}$ and calculate $D(R^{{(n)}}_{\rm sum},\textbf{x}^{(n)})$.
                \While {$norm(D(R^{{(n)}}_{\rm sum},\textbf{x}^{(n)}))\geq tolerance$}
                \State Update $\textbf{x}^{(n)}$: $\textbf{x}^{(n)}=\textbf{x}^{(n)}+\alpha D(R^{{(n)}}_{\rm sum},\textbf{x}^{(n)})$.
                \State Construct $\textbf{A}^{(n)}$ from $\textbf{x}^{(n)}$.
                \State Scale $\textbf{A}^{(n)}$ and construct $\textbf{x}^{(n)}$.
                \State Calculate $D(R^{{(n)}}_{\rm sum},\textbf{x}^{(n)})$.
                \EndWhile
                \EndFor
                \State Calculate $\textbf{G}^{(1)},\cdots,\textbf{G}^{(N-1)}$ using (\ref{GG}).
        \end{algorithmic}
\end{algorithm}
\vspace{-1em}
 

\subsection{Convergence Behavior and Computational Complexity Analysis}\label{complexity}

In this section, we discuss the convergence behavior of our proposed algorithms and analyze their computational complexity in comparison with the existing schemes.

The number of iterations needed for the proposed optimizations depends on the step size $\alpha$, and there is a natural trade-off between the convergence rate and the achieved sum-rate. Here, we simply choose $\alpha=0.03$ for the separate optimization and $\alpha=0.01$ for the joint optimization based on experience. By stopping the iterations when less than 5\% improvement is observed over one iteration, the separate optimization algorithm converges after around 75 iterations and the joint one converges after about 100 iterations.

Next, we analyze the computational complexities of the proposed joint and separate optimization algorithms for our PZF beamforming design, and compare them with those of ZF, MMSE, RZF, and MF beamforming schemes in \cite{amah2011non1,amah2010beamforming}, and \cite{peel2005vector}, as well as the scheme proposed in \cite{li2017joint}. The order of complexity with respect to the number of relay antennas $M$, number of users $N$, and the iteration number $iter$, is used for the analysis.

 The mathematical operations in the beamforming matrix optimization include summation, multiplication, division, square root, sorting, taking logarithm, and comparison. Among these operations, division and multiplication have the highest computational complexity and the highest numbers of happening, while other operations lead to much lower computational complexity. So, our analysis focuses on division and multiplication. The required numbers for each of the two operations in each beamforming design are listed in TABLE \ref{complexity}.
\begin{table}[H]
	
	\centering
	\small
	\begin{tabular}{|c| c| c|}
		
		\hline
		{Scheme$\backslash$}{Operations}
		&
		$\times$ &
		$\div$  \\
		
		\hline		
		
		{ZF}&  {$N^3(2M)$}   & {$2N^2+\dfrac{M^2}{2}$}\\
		\hline

		{MMSE and RZF} & {$N^3(2M)$}&  {$M^2$} \\
		\hline		
		
		{MF}&  {$N^3(2M)$}& {$\dfrac{M^2}{2}$} \\
		\hline		
		
		{PZF-Joint}&  {$iter \times N^5(6M^2)$ }  & {$iter \times 2N^5$}\\
		\hline

		{PZF-Separate}&  {$iter \times N^4(6M^2)$}&  { $iter \times 2N^4$} \\
		\hline		
		
	\end{tabular}
	
	\caption{The numbers of multiplications and divisions in the design of PZF (joint and separate), ZF, MMSE, RZF, and MF schemes.}

	\label{complexity}
\end{table}

As can be seen from TABLE \ref{complexity}, the numbers of multiplications for ZF, MMSE, RZF and MF schemes are the same. Actually, this value comes from the calculation of $\textbf{G}^{(n)}=\textbf{G}_{\text{TX}}^{(n)}\textbf{P}^{n}\textbf{G}_{\text{RX}}$.  For both proposed PZF beamforming methods, the dominant parts for the multiplications are resulted from the calculations of $|\textbf{h}_k^T \textbf{G}^{(n)} \textbf{h}_i|^2$ and $|\textbf{h}_k^T \textbf{G}^{(n)}|^2$, while for the divisions they are resulted from the calculations of ${P_i|\textbf{h}^T_k\textbf{G}^{(n)}\textbf{h}_i|}^2/({|\textbf{h}_k^T\textbf{G}^{(n)}|}^2+1)$.

        %
        %
        %
        %
        %
        %
        %
        %
        %
        %
        
        TABLE \ref{complexity} also shows that both proposed beamforming optimizations bear higher computational complexity than other common schemes. This higher complexity is due to the iterative feature of our algorithms, and the fact that we optimize more elements in the transformations of the beamforming matrices, i.e., our optimization problems have higher dimensions. For the joint algorithm, the whole beamforming matrices for different time slots are optimized together which leads to higher complexity than the separate algorithm. As discussed earlier in this subsection, the iteration number is about 100 for the joint optimization and 75 for the separate optimization. Another insight from this table is that for large number of antennas, i.e., when $M$ is large, the complexity of our PZF beamforming is still tractable, as it is only one order of magnitude higher than the other schemes. On the other hand, for large numbers of antennas and users, i.e., when both $M$ and $N$ are large, the complexity increases 3 orders of magnitude faster than the other schemes which may make it intractable. 

Recently, another beamforming strategy is proposed in \cite{li2017joint}, where the relay beamforming and receiver processing  matrices  are jointly designed and multi-symbol processing is used at each user. While our work targets sum-rate optimization, \cite{li2017joint} studied the SINR max-min optimization. The computational complexity of the scheme in \cite{li2017joint}, is $O(iter\times N^4 M^6)$, which has a higher order than both of our schemes. Further, it has a higher decoding complexity of $O(N^2)$ at the users due to the multi-symbol processing. 
In TABLE I of \cite{li2017joint}, the authors have shown a comparison between the average CPU processing time for their approach and our approach when  $N=3, M=3$, and $N=4, M=4$, which declares that ours is about 5 times faster.

The proposed algorithms for partial zero-forcing are based on gradient-ascent method and there is no global convergence guarantee. Since ZF relay matrices are selected as the initialization point, the solutions found by our algorithms are guaranteed to achieve higher achievable sum-rates than ZF. To see this, we show that ZF relay matrices are not local optima. We first consider the joint optimization and look at the problem given in (\ref{f_jontopt})-(\ref{const_jointopt}).  To show that ZF relay matrices are not local optima of the optimization problem,  it is sufficient to prove that $D(f,\textbf{x}_{\rm ZF})= 0$
does not hold, where $\textbf{x}_{\rm ZF}$ is the  corresponding vector of the ZF relay matrices, ${\textbf{G}_{\text{ZF}}}^{(n)}$ for $n = 1,...,N-1$. Notice that $D(f,\textbf{x}_{\rm ZF})$ is a function of the channel matrix $\bf H$. From (\ref{RS-E}) and (\ref{AA}), it can easily be shown that 
${\bf A}_{\rm  ZF}^{(n)}=\frac{1}{p_{\rm ZF}^{(n)}}{\bf \textbf{P}}^{n}$, thus according to (\ref{A-set}), (\ref{xxn1}), and (\ref{xxn2}), $\textbf{x}_{\rm ZF}$ has $(N-1)N$ non-zero coefficients\footnote{For instance, for the case $N=3$, $\textbf{x}^{(1)}_{\rm ZF}=\frac{1}{p_{\rm ZF}^{(1)}}[0\quad 1\quad 0\quad 1\quad 1\quad 0]$, and $\textbf{x}^{(2)}_{\rm ZF}=\frac{1}{p_{\rm ZF}^{(2)}}[0\quad 0\quad 1\quad
1\quad 0\quad 0\quad 0\quad 1\quad 0]$.}, as there are $N$ non-zero elements in each ${\textbf{P}^{n}}$.   From  (\ref{NumOpt}), the number of equations in $D(f,\textbf{x}_{\rm ZF})=0$ is $W=(N+2)N(N-1)/2$, which is larger than the number of non-zero parameters in ${\textbf x}_{\rm ZF}$. As the $M \times N$ channel matrix $\bf H$ is random whose entries are i.i.d.~following Rayleigh distribution, the probability that $D(f,\textbf{x}_{\rm ZF})=0$ holds, is zero even when the non-zero coefficients are adjustable. This means that with probability 1, ZF relay matrices are not local optima of the optimization problem. Via similar reasoning, it can be shown that the ZF beamforming matrices cannot be the local optima for the separate optimization method as well. Simulations have also suggested this conclusion and we always obtain higher sum-rates than the ZF beamforming.

\vspace{-.5em}
\subsection{Simulation Results}
        \label{sec4.1}
        In this section, we show simulation results on the sum-rate of MWRNs with our PZF design and other existing designs \cite{amah2011non1,amah2010beamforming,li2017joint}. We choose $M=N=3$, i.e., 3 single-antenna users communicate with each other with the help of a relay equipped with 3 antennas. 
        
\begin{figure}[!ht]
             \centering
             \includegraphics[width=3.6in]{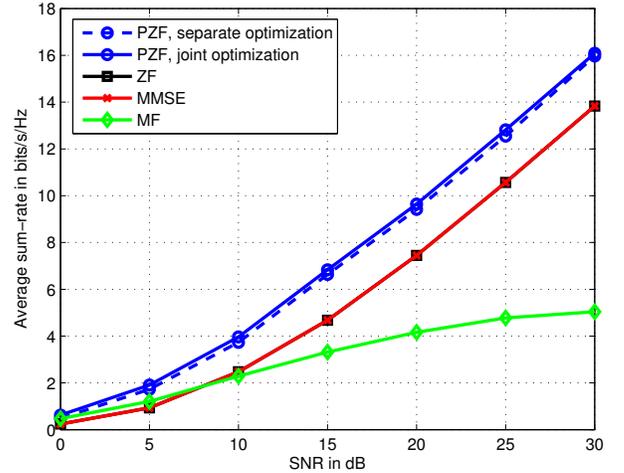}
             \caption{Sum-rates for a homogeneous 3-user MWRN with $P_R=1$.}
             \label{SRr2}
             \vspace{-1em}
        \end{figure}

        First, we consider a homogeneous network where all the channels follow i.i.d. $\mathcal{CN}(0,\sigma_h^2)$. We set $P_R=P_1=P_2=P_3=1$, thus the SNR of each user at the relay will be $\sigma_h^2$. Figure \ref{SRr2} shows the sum-rates for different SNR values. We can see that the proposed PZF design has the best sum-rate performance for the whole SNR range. It can also be observed that for the proposed PZF scheme, the separate and joint optimization methods give very close sum-rate performances with the latter slightly better. Simulation results on the sum-rate comparison between our proposed scheme and the one in \cite{li2017joint} is available in Figures 13 and 14 of \cite{li2017joint}, for the cases of $N=3, M=3$, and $N=4, M=4$. It can be observed that our proposed joint design has slightly lower sum-rate performance and the gap shrinks as the SNR increases.

        \begin{figure}[!ht]
                \centering
                \includegraphics[width=3.6in]{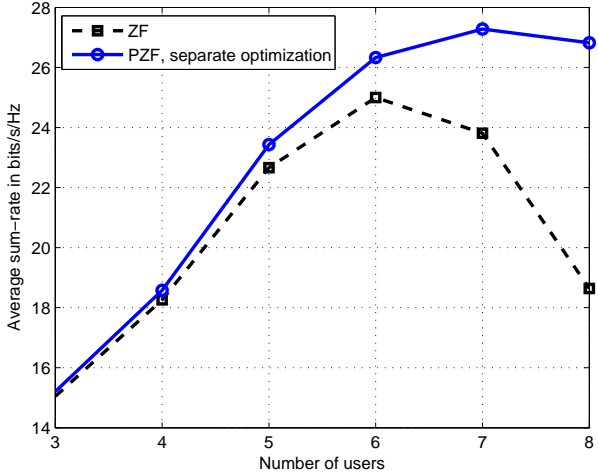}
                \caption{Sum-rates of PZF and ZF schemes with unicasting strategy for different numbers of users, $M=8$, and SNR$=$20 dB.}
                \label{MAtennaN}
                \vspace{-1em}
        \end{figure}
        
        Moreover, we have presented simulation results on the sum-rates of MWRNs with different numbers of users. We consider that the MWRN has an 8-antenna relay, and the number of users changes from 3 to 8. We set $P_R=P_1=P_2=\cdots=P_8=1$ and SNR = $20$ dB. The channels follow i.i.d. $\mathcal{CN}(0,\sigma_h^2)$. Figure \ref{MAtennaN} shows the relationship between the number of users and the sum-rate for the ZF and PZF schemes. From the figure, we can conclude that the network sum-rate first increases and then decreases as the number of users increases. Also, it can be seen that the advantage of PZF over the ZF design enlarges with more number of users. The reason for this is two-fold. First, compared to ZF,  PZF beamforming allows extra $N(n+1)$ degrees-of-freedom in the design of $\textbf{G}^{(n)}$. So, as the number of users $N$ increases, there are more extra degrees-of-freedom in the PZF design compared to ZF. Another contributing factor is the ZF beamforming coefficient, $\frac{1}{p_{\text{ZF}}^{(n)}}$, which tends to decrease when $N$ increases. This  leads to a lower SNR and thus a lower achievable sum-rate.

         \begin{figure}[!ht]
				\centering
         		\includegraphics[width=3.6in]{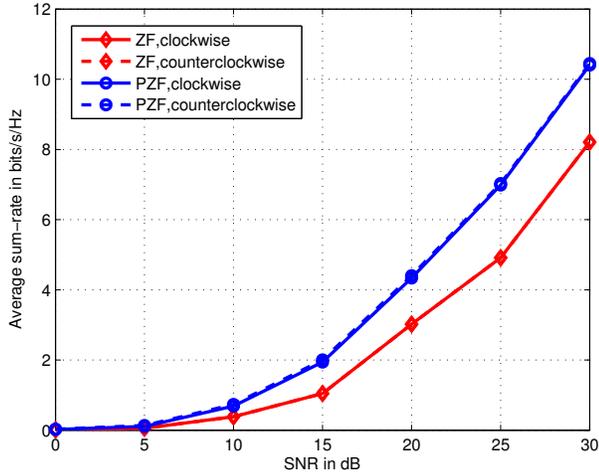}
	          	\caption{Sum-rates for a heterogeneous 3-user MWRN with $P_R=1$, through separate optimization.}
	         	\label{NonId}
	         	\vspace{-.5em}
         \end{figure}
         
         In addition, we consider a 3-user MWRN with non-identical fading channels due to different path-losses. Denote $d_i$ as the distance from an arbitrary user, $u_i$, to the relay. The channels between $u_i$ and the $M$ relay antennas, $h_{m,i}$s, are assumed to follow $\mathcal{CN}(0,\sigma_i^2)$, where $\sigma^2_{i}=(\psi /d_i)^{\nu}$ with $\psi$ being a constant. In simulations, we set $d_3=2d_2=4d_1$ and assume $\nu=2$.  With this heterogeneous setup, the decoding order may affect the sum-rate, thus we consider 2 orders of detection: clockwise as defined in \eqref{relation}, and counter clockwise defined as $i=\text{mod}_N(k-n-1)+1$. In Figure \ref{NonId}, the x-axis, denoted as SNR, shows $u_1$'s SNR at the relay, thus $\text{SNR}=\sigma_{1}^2= {4}\sigma_2^2={16}\sigma_3^2$. We can see from this figure that the proposed PZF design achieves a significantly higher sum-rate than the ZF design. For both clockwise and counter clockwise detections, ZF provides exactly the same sum-rate performances, while PZF provides slightly different performances. For systems with more users or relay antennas, the advantage of adopting a better decoding order may become larger as the difference between the channel qualities of different users will become larger on average. So, in this case the issue of finding the optimal decoding order becomes more important.  Since our focus is the new PZF relay beamforming design not decoding order, we refer further investigations on how the decoding order affects the PZF scheme to future work.
         
         \begin{figure}[!ht]
         	\centering
         	\includegraphics[width=3.6in]{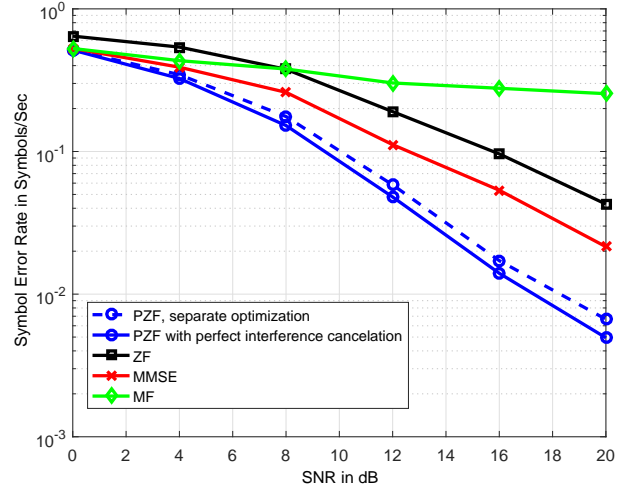}
         	\caption{SERs for a homogeneous 3-user MWRN with $P_R=1$.}
         	\label{fig:6}
         	\vspace{-.5em}
         \end{figure}
         
\begin{figure}[!ht]
		 	\centering
		 	\includegraphics[width=3.6in]{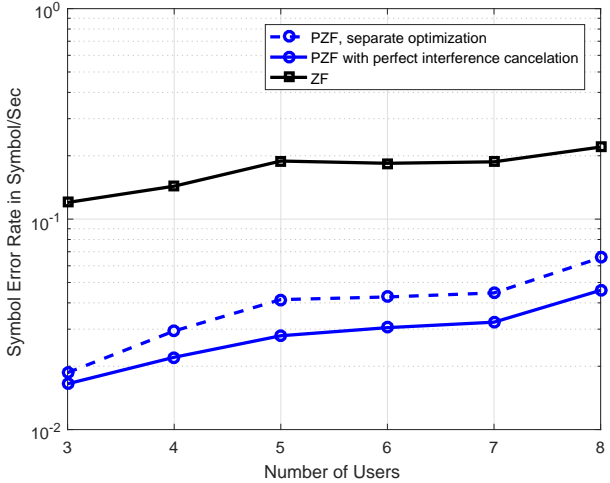}
		 	\caption{SER versus the number of users for homogeneous networks, where $M=N$ and SNR=15 dB.}
		 	\label{SER_N}
		 	\vspace{-1.5em}
			 \end{figure}

		Next, we compare the symbol error rate (SER) of PZF beamforming with the ones of ZF, MMSE, and MF schemes. The SER results for MWRNs with $M=N=3$ are shown in Figure \ref{fig:6}, where quadrature amplitude modulation (QAM) is used for all users' symbols. It can be seen that our PZF design provides a far lower SER than ZF,  MF, and MMSE schemes. Furthermore, to see the effect of error propagation, which is the detection error of a symbol caused by the symbol detection errors in previous time slots, we have presented the simulation results for the ideal case of perfect interference cancellation for the PZF scheme. In this ideal scheme, for every BC time slot, we cancel the interference caused by the previously decoded signals using the correct and error-free symbols instead of using the decoding results from the previous time slots. This way, no decoding errors in previous BC time slots can propagate to the coming BC time slots. The simulations show that the effect of error propagation is negligible for our proposed PZF scheme. Moreover, in order to see the behavior of error propagation when the number of users increases in homogeneous networks, we have presented the simulation results for SER versus the number of users, where $M=N$ changes from 3 to 8. As it is shown in Figure \ref{SER_N}, the effect of error propagation slightly increases as the user number increases, but it is still very small in comparison to the performance enhancement that our proposed beamforming has brought compared to the ZF scheme.

			 \begin{figure}[!ht]
			 	\centering
			 	\includegraphics[width=3.6in]{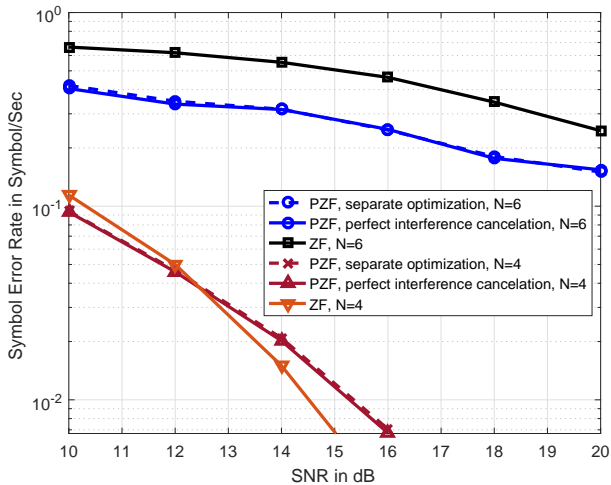}
			 	\caption{SER performance of heterogeneous networks for $N=4$ and $N=6$ cases when $M=32$ with $P_R=1$.}
			 	\label{SER_heter}
			 	\vspace{-.5em}
			 \end{figure}
In addition to homogeneous networks, in Figure~\ref{SER_heter}, we have presented the results of SER versus SNR of heterogeneous networks for $N=4$ and $N=6$ cases when $M=32$ antennas are available at the relay. Recall that the channels between an arbitrary user, $u_i$, and the relay antennas, $h_{m,i}$s, follow $\mathcal{CN}(0,\sigma_i^2)$, where $\sigma^2_{i}=(\psi/d_i)^{\nu}$ with $\psi$ being a constant, and $d_i$ being the distance from $u_i$ to the relay. In this simulation, we have set $d_n=2^{(n-1)}d_1$ for $n=1, 2, ..., N$ and $\nu=2$. The SNR in Figure~\ref{SER_heter} represents the SNR of the first user, $u_1$, at the relay.  This figure shows the results for the clockwise order of decoding. It can be seen from Figure~\ref{SER_heter} that the effect of error propagation diminishes as SNR grows, also we have more error propagations when there are higher number of users. Moreover, it can be observed that PZF gives better SER performances than ZF for $N=6$, while when $N=4$ this may not be the case in higher SNR ranges. The reason for this is two-fold. 1) Our optimization targets at sum-rate maximization not SER optimization, which may lead to degraded SER performance. 2) The number of degrees-of-freedom increases as the number of users increase, so we can achieve better SER performances in comparison to ZF when higher number of users are involved.

        \section{PZF with Hybrid Uni/Multicasting}
        \label{section4}
        
        In Section \ref{sec PZF}, we considered that in each BC time slot, the relay transmits uniquely different information symbols to different users, which is called transmission via unicasting. However, in this section, the hybrid uni/multicasting strategy is considered. It is shown that when the relay uses uni/multicasting strategy, PZF is still able to improve the sum-rate performance of MWRNs.
        
        \subsection{Hybrid Uni/Multicasting Strategy}
        Along with the unicasting strategy, hybrid uni/multicasting is also proposed in \cite{amah2011regenerative}. If hybrid uni/multicasting strategy is used, in each BC time slot, one information symbol is exclusively transmitted to one user (unicast transmission), and another information symbol is transmitted to the other $N-1$ users (multicast transmission). The unicasted information symbol is fixed in all BC time slots, and transmitted to different users in different BC time slots. While the multicasted information symbols are changed in different BC time slots. This hybrid uni/multicasting scheme ensures that each user receives all other users' symbols within the $N-1$ BC time slots. Detection scheduling of this transmission strategy will be discussed in Subsection \ref{Detection}.
        \begin{figure}[!ht]
                \centering
                \includegraphics[width=.34\textwidth]{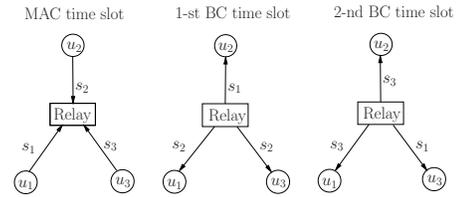}
                \caption{Hybrid uni/multicasting strategy.}
                \label{15b} 
				\vspace{-1em}
        \end{figure}
        
        A 3-user example of hybrid uni/multicasting strategy is shown in Figure \ref{15b}. In the MAC phase, simultaneously, $u_1$ sends $s_1$, $u_2$ sends $s_2$, and $u_3$ sends $s_3$  to the relay. In the BC phase, $s_1$ is chosen as the unicasting symbol, while $s_2$ and $s_3$ are chosen as the multicasting symbols for the first and second time slot, respectively. In the first BC time slot, $u_1$ and $u_3$ decode $s_2$ and $u_2$ decodes $s_1$, from the relay broadcast signal. In the second BC time slot, $u_1$ and $u_2$ decode $s_3$,  and $u_3$ decodes $s_1$. After the MAC and BC phases, each user decodes the information symbols from all other users.
        
        PZF can naturally be extended to the hybrid uni/multicasting transmission strategy. The only modification in the problem formulation of sum-rate maximization that needs to be done, is the structures of $\textbf{A}^{(n)}$ matrices (or the locations of the zero entries in $\textbf{A}^{(n)}$s) which should be adjusted based on the hybrid uni/multicasting strategy. For example, for the aforementioned 3-user network, $\textbf{A}^{(n)}$ matrices should have the following forms
        \begin{equation}
        \textbf{A}^{(1)}_{\text{PZF}}=
        \left(                
        \begin{array}{ccc}   
        * & * & 0\\  
        * & * & 0  \\ 
        0 & * & *
        \end{array}
        \right),
        \textbf{A}^{(2)}_{\text{PZF}}=
        \left(                
        \begin{array}{ccc}   
        * & * & *\\  
        * & * & *  \\ 
        * & * & *
        \end{array}
        \right).
        \end{equation}
         So, this problem can be solved by the modified gradient-ascent method proposed in Section \ref{sec PZF}.  
        
        \subsection{Simulation Results}
        
\begin{figure}[!ht]
                \centering
                \includegraphics[width=3.6in]{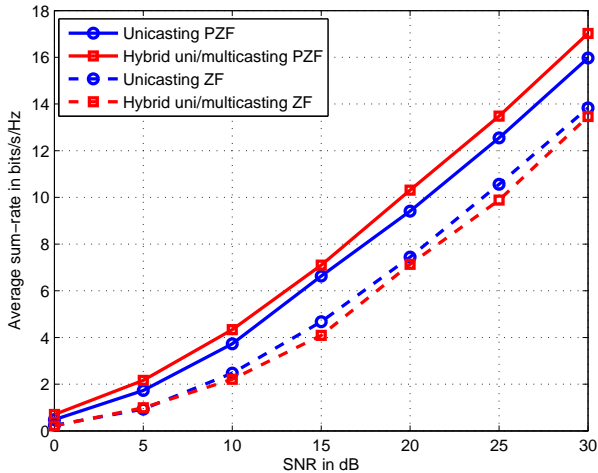}
                \caption{Sum-rates of homogeneous 3-user MWRNs with unicasting and hybrid uni/multicasting transmission strategies for ZF beamforming and PZF design through separate optimization.}
                \label{UniMulti}  
				\vspace{-1.5em}
	        \end{figure}

                This section shows the simulation results on the sum-rate of MWRNs with our proposed PZF scheme, and compares it with ZF beamforming design when hybrid uni/multicasting is the transmission strategy at the relay. $M=N=3$ is chosen, i.e., 3 single-antenna users communicate with each other with the help of a relay equipped with 3 antennas. We consider a homogeneous network where all channels follow i.i.d. $\mathcal{CN}(0,\sigma_h^2)$, and we set $P_R=P_1=P_2=P_3=1$.  Figure \ref{UniMulti} shows the sum-rates for different SNR values. We can see that the hybrid uni/multicasting PZF design has a better sum-rate performance than the hybrid uni/multicasting ZF design for the whole SNR range. It can also be observed that when hybrid uni/multicasting is used, the sum-rate performance gap between PZF and ZF designs becomes larger.
        
        \subsection{Discussion on Detection Scheduling}\label{Detection}
\begin{figure}[!ht]
                \centering
                \includegraphics[width=.34\textwidth]{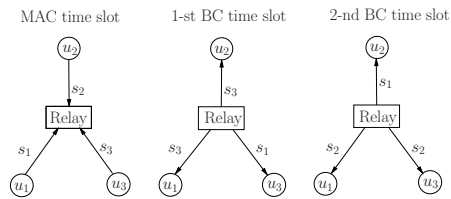}
                \caption{Another scheduling of detection for hybrid uni/multicasting strategy.}
                \label{zf2} 
                \vspace{-.5em}
        \end{figure}

        Unlike the unicast model, hybrid uni/multicast strategy brings imbalance in the transmission of different users' symbols. For MWRNs with asymmetric channel conditions, the choices of signals for unicasting and multicasting in different BC time slots, or the scheduling of detections in the BC phase may affect the sum-rate performance. Commonly, the channel condition is used to decide which user's signal should be unicasted and the order of other users' signals to be multicasted in the BC time slots. It is beneficial to multicast the signals of users with good channel conditions during earlier BC time slots, and the ones with poor channel conditions during latter BC time slots. The choice of  users'  signals to be unicasted is complicated and needs further study.
In Figure \ref{15b}, $s_2$ and $s_3$ are chosen to be the multicasted symbols in the first and second BC time slots, respectively. Consequently, $u_1$ first decodes $s_2$ and then $s_3$. A different scheduling of detection is shown in Figure \ref{zf2}, where $s_3$ is multicasted in the first BC time slot, and $s_2$ is multicasted in the second BC time slot. Correspondingly, $u_1$ first decodes $s_3$ and then $s_2$. 
        
\begin{figure}[!ht]
                \centering
                \includegraphics[width=3.6in]{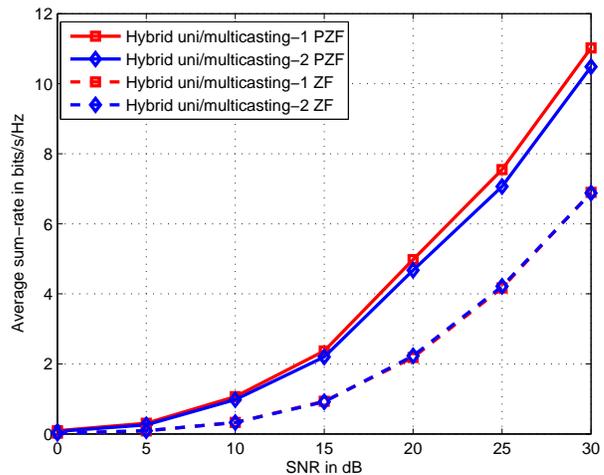}
                \caption{Sum-rates of hybrid uni/multicasting strategy using different decoding schedules in a 3-user MWRN  with ZF beamforming and PZF design through separate optimization.}
                \vspace{-1.5em}
                \label{schedule}  
        \end{figure}

        Simulation results on different kinds of  detection scheduling are given in Figure  \ref{schedule}. We consider a 3-user MWRN with non-identical fading channels due to different path-losses. The same as Subsection \ref{sec4.1}, we denote $d_i$ as the distance from an arbitrary user, $u_i$, to the relay. The channels between $u_i$ and the $M$ relay antennas, $h_{m,i}$s, are assumed to follow $\mathcal{CN}(0,\sigma_i^2)$, where $\sigma^2_{i}=(\psi /d_i)^{\nu}$ with $\psi$ being a constant. In simulations, we set $d_3=2d_2=2d_1$, and assume $\nu=2$. With this heterogeneous setup, we consider 2 kinds of  detection scheduling: one is described in Figure \ref{15b}, denoted as hybrid uni/multicasting-1, and the other is shown in Figure \ref{zf2}, denoted as hybrid uni/multicasting-2.
        In Figure \ref{schedule}, the x-axis, denoted as SNR, shows $u_1$'s SNR at the relay, thus $\text{SNR}=\sigma_{1}^2=\sigma_2^2={4}\sigma_3^2$. From this figure, we can see that when PZF is applied at the relay, hybrid uni/multicasting-1 has higher sum-rates than hybrid uni/multicasting-2. An explanation for this observation is that in the simulation settings, the channels between $u_3$ and the relay are weaker than the channels between the other two users and the relay. Thus, as hybrid uni/multicasting-1 chooses to decode the weakest signal, $s_3$, in the last BC time slot, it leads to higher sum-rates. 
        
        \section{PZF for MWRNs Where $M=N-1$}\label{secV}
        In ZF beamforming, the number of relay antennas, $M$, must be larger than, or at least equal to the number of users, $N$, i.e., $M\ge N$. Otherwise, there will not be enough degrees-of-freedom to remove users' interferences \cite{jafar2008degrees}. However, in PZF beamforming, due to the fact that the interferences do not need to be fully canceled, the number of antennas at the relay can be reduced by one. In other words, PZF can be used for MWRNs where $M\ge N-1$. The case of $M \geq N$ has been considered in Sections \ref{sec PZF} and \ref{section4}. In this section, we consider MWRNs where the number of relay antennas is one less than the number of users, i.e, $M=N-1$. The transceiver protocol is the same as  Sections \ref{sec PZF} and \ref{section4}. So, there are one MAC time slot and $N-1$ BC time slots for the multi-way communications. Also, with PZF, in each BC time slot for each user, only partial interference (interference excluding self-interference and interferences from previously decoded signals) needs to be canceled. However, the problem formulation of PZF relay beamforming design for the $M=N-1$ case, is largely different from the one in Section \ref{sec PZF}. In fact, as the number of relay antennas is smaller than the number of users, the dimension of $\textbf{G}^{(n)}$ is smaller than the dimension of $\textbf{A}^{(n)}$, and thus, the map from $\textbf{A}^{(n)}$ to $\textbf{G}^{(n)}$, in \eqref{GG}, does not apply. As a result, the sum-rate optimization needs to be conducted with respect to $\textbf{G}^{(n)}$, directly.  This way the optimization problem formulation will be as follows:
        \begin{eqnarray}\label{2JointOpt0}
        &&\hspace{-18mm}\max_{\textbf{G}^{(1)},\cdots,\textbf{G}^{(N-1)}}\hspace{-1mm}\sum^{N}_{i=1}\min_{k \neq i}\left\{\log_{2}\Bigg(1\hspace{-1mm}+\hspace{-1mm}\frac{{P_i|\textbf{h}^T_k\textbf{G}^{(n)}\textbf{h}_i|}^2}{{|\textbf{h}_k^T\textbf{G}^{(n)}|}^2\hspace{-1mm}+1}\Bigg)\hspace{-1mm} \right\} \label{2J-opt10}\\
        \hspace{-5mm}&\text{s.t.}& {\rm tr}\left\{\textbf{G}^{(n)}\left(\textbf{H}\textbf{P}_\textbf{s}\textbf{H}^{H}+\textbf{I}\right)(\textbf{G}^{(n)})^{H}\right\}\le P_{R}, \label{2J-opt20}\\
        &\mbox{and}& \textbf{H}^T\textbf{G}^{(n)}\textbf{H}=\textbf{A}^{(n)}, \mbox{ for } n=1,2,\cdots, N-1. \label{2nonCon0}
        \label{2opt-prob0}
        \end{eqnarray}
        
        According to \eqref{NumNonzero}, there are $(N-n-1)N$ zero-valued entries in $\textbf{A}^{(n)}$ and the rest of the entries can take any complex value. Equation \eqref{2nonCon0} can be written as $(N-n-1)N$ linear homogeneous equations. So to simplify the optimization problem we define vector $\textbf{g}^{(n)}$ that contains all entries in $\textbf{G}^{(n)}$, as
        \begin{equation}
        \textbf{g}^{(n)}=[g^{(n)}_{11}\ \ g^{(n)}_{12}\ \ \cdots \ \ g^{(n)}_{1,N-1} \ \ g^{(n)}_{21}\ \ \cdots \ \ g^{(n)}_{N-1,N-1}].
        \label{Gxxn2}
        \end{equation}
        Then, we divide $\textbf{g}^{(n)}$ into two vectors, $\textbf{y}^{(n)}$ and $\textbf{r}^{(n)}$, where $\textbf{y}^{(n)}$ contains the first $(N-1)^2-(N-n-1)N$ entries of $\textbf{g}^{(n)}$, and $\textbf{r}^{(n)}$ contains the rest $(N-n-1)N$ entries. Since the number of entries in $\textbf{r}^{(n)}$ is equal to the number of linear equations in \eqref{2nonCon0}, $\textbf{r}^{(n)}$ can be uniquely represented by $\textbf{y}^{(n)}$ from \eqref{2nonCon0}. 
        This way the constraints in \eqref{2nonCon0} will be eliminated.
        
        Based on the above discussion, the sum-rate maximization problem is transformed into an optimization over $\textbf{y}^{(n)}$, with the only constraint in \eqref{2J-opt20}. Thus, the proposed modified gradient-ascent method can be used. The detailed algorithm is given in Algorithm \ref{2SepAlgorithm}, where for the complexity considerations separate optimization of the relay beamforming matrices is considered.
        
        \begin{algorithm}[!ht]
                \caption{Separate optimization scheme for MWRNs where $M=N-1$.}\label{2SepAlgorithm}
                \begin{algorithmic}[1]
                        \State Initialize $\alpha$ and $tolerance$.
                        \For {$n=1:N-1$}
                        \State Initialize $\textbf{y}^{(n)}$ and construct $\textbf{G}^{(n)}$ by solving \eqref{2nonCon0}.
                        \State Scale $\textbf{G}^{(n)}$ to satisfy \eqref{2J-opt20} and construct $\textbf{y}^{(n)}$.
                        \State Calculate $D(R^{{(n)}}_{\rm sum},\textbf{y}^{(n)})$.
                        \While {$norm(D(R^{{(n)}}_{\rm sum},\textbf{y}^{(n)}))\geq tolerance$}
                        \State Update $\textbf{y}^{(n)}$: $\textbf{y}^{(n)}=\textbf{y}^{(n)}+\alpha D(R^{{(n)}}_{\rm sum},\textbf{y}^{(n)})$.
                        \State Construct $\textbf{G}^{(n)}$ from $\textbf{y}^{(n)}$ by solving \eqref{2nonCon0}.
                        \State Scale $\textbf{G}^{(n)}$ to satisfy \eqref{2J-opt20} and construct $\textbf{y}^{(n)}$.
                        \EndWhile
                        \EndFor
                \end{algorithmic}
                 \end{algorithm}

        \begin{figure}[!ht]
				\centering
                \includegraphics[width=3.6in]{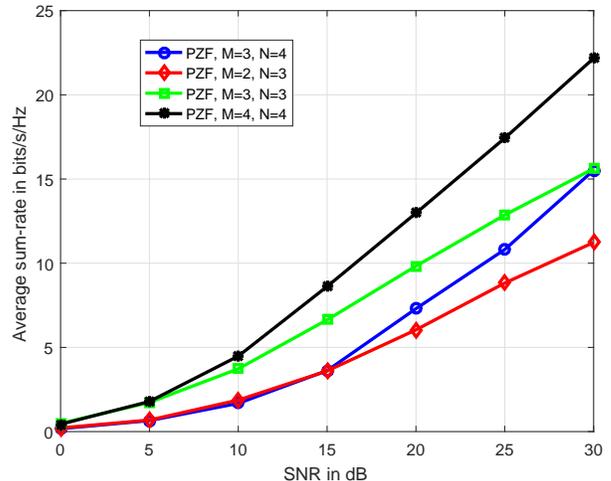}
                \caption{Sum-rates of PZF design with separate optimization and unicasting strategy for four network settings when $M=N$ and $M=N-1$.}
                \label{sum_rate_Mantennas}
				\vspace{-1.5em}
        \end{figure}
        
        Next, we show the simulation results on the sum-rates of MWRNs, where $M=N-1$ and  our PZF beamforming design with unicasting is applied. We set $P_R=P_1=P_2=P_3=P_4=1$ and the channels are considered to be homogeneous and follow i.i.d. $\mathcal{CN}(0,\sigma_h^2)$. Figure \ref{sum_rate_Mantennas} shows the sum-rates of PZF design for two cases, when the relay has 2 and 3 antennas, and the number of users is 3 and 4, respectively. It can be seen from this figure that 1) the sum-rates for $M=N=4$ are the highest, 2) the $M=N=3$ case achieves higher sum-rates than the case of $M=2, N=3$, and 3) compared to the case of $M=3, N=4$, the sum-rates of the $M=N=3$ case are higher for the SNR range of [0 dB,30 dB], but the advantage decreases with SNR and the curves indicate that the case of $M=3, N=4$ outperforms the case of $M=N=3$ when SNR is higher than 30 dB. Further, we can observe that in the high SNR regime, the case of $M=3, N=4$ gives higher sum-rates than the case of $M=2, N=3$. However, in low SNR regime, the sum-rates are similar.

\section{Conclusion}
In this paper, a novel PZF relay beamforming design is proposed for MWRNs where $N$ single-antenna users communicate with each other with the help of one $M$-antenna relay. Compared with ZF relay beamforming, the proposed scheme allows more degrees-of-freedom in the beamforming optimization and thus, can improve the sum-rate. On the other hand, with the help of self-interference cancellation and successive interference cancellation, the proposed design enables interference-free communications.

 For the case when the number of users is no larger than the number of relay antennas, design of the PZF relay beamforming matrices was firstly transformed into the design of equivalent channel matrices. Then a modified gradient-ascent method was proposed to solve the optimization problems both jointly and separately. The convergence behavior of the proposed algorithms was studied, and computational complexity comparison was provided between the proposed methods and the existing ones. Simulations on the achievable sum-rate and symbol error rate have shown that significant performance improvement is obtained  with the proposed new designs. Further, extensions of the proposed schemes are made to MWRNs with hybrid uni/multicasting transmission strategy and MWRNs where the number of users is one more than the number of relay antennas. Similar advantages have been achieved with the proposed PZF idea in these cases.


%

\ifCLASSOPTIONcaptionsoff
  \newpage
\fi

\bibliographystyle{IEEEtran}
\bibliography{reference}

\end{document}